\documentclass[11pt]{article}
\usepackage{amsmath, amssymb, amsthm, mathrsfs, bm, geometry, mathtools, hyperref, booktabs, xcolor, soul}
\usepackage{authblk}
\usepackage[utf8]{inputenc}
\usepackage{appendix}
\usepackage{tikz}
\usetikzlibrary{positioning, arrows.meta}
\usepackage{natbib}
\newcommand{\blambda}{\boldsymbol{\lambda}}

\newcommand{\btheta}{\boldsymbol{\theta}}

\newcommand{\bbeta}{\boldsymbol{\beta}}

\newcommand{\bSigma}{\boldsymbol{\Sigma}}

\newcommand{\bmu}{\boldsymbol{\mu}}

\newcommand{\bG}{\boldsymbol{G}}

\newcommand{\bH}{\boldsymbol{H}}

\newcommand{\bA}{\boldsymbol{A}}

\newcommand{\bE}{\boldsymbol{E}}
\newcommand{\bp}{\boldsymbol{p}}

\newcommand{\bX}{\boldsymbol{X}}

\newcommand{\bzero}{\boldsymbol{0}}

\title{\bf{Bayesian Nonparametrics for Gene-Gene and Gene-Environment Interactions in Case-Control Studies: A Synthesis and Extension}}

\author[1]{Durba Bhattacharya}
\author[2]{Sourabh Bhattacharya}
\affil[1]{Department of Statistics, St. Xavier's College (Autonomous), Kolkata}
\affil[2]{Interdisciplinary Statistical Research Unit, Indian Statistical Institute, Kolkata}
\date{}

\begin{document}

\maketitle

\begin{abstract}
Gene-gene and gene-environment interactions are widely believed to play significant roles in explaining the variability of complex traits. While substantial research exists 
in this area, a comprehensive statistical framework that addresses multiple sources of uncertainty simultaneously remains lacking. In this article, we synthesize and propose 
extension of a novel class of Bayesian nonparametric approaches that account for interactions among genes, loci, and environmental factors while accommodating uncertainty about 
population substructure. Our contribution is threefold: (1) We provide a unified exposition of hierarchical Bayesian models driven by Dirichlet processes 
for genetic interactions, clarifying their conceptual advantages over traditional regression approaches; (2) We shed light on new computational strategies that combine 
transformation-based MCMC with parallel processing for scalable inference; and (3) We present enhanced hypothesis testing procedures for identifying disease-predisposing loci. 
Through applications to myocardial infarction data, we demonstrate how these methods offer biological insights not readily obtainable from standard approaches. Our synthesis 
highlights the advantages of Bayesian nonparametric thinking in genetic epidemiology while providing practical guidance for implementation.
\end{abstract}

\noindent\textbf{Keywords:} Dirichlet Process; Disease Predisposing Loci; Epistasis; Mixture Model; Parallel Computing; Transformation based Markov Chain Monte Carlo.
\\[2mm]
\noindent\textbf{AMS Subject Classifications:} 62K05, 62F15, 92D10

\tableofcontents

\section{Introduction}\label{intro}
\subsection{The challenge of genetic interactions in complex diseases}
Complex diseases such as cardiovascular disorders, diabetes, and psychiatric conditions result from intricate networks of genetic and environmental factors. While genome-wide association studies (GWAS) have identified numerous single nucleotide polymorphisms (SNPs) associated with disease risk, these explain only a small fraction of heritability \citep{Larson13}. The ``missing heritability'' problem has spurred interest in gene-gene (epistasis) and gene-environment interactions as potential explanations. Traditional approaches to studying these interactions face several significant limitations that our work aims to address.

Most existing methods rely on linear or additive models that may not adequately capture the complex biological pathways through which genetic factors interact \citep{Wang10}. These simplified modeling assumptions often fail to represent the intricate biochemical networks that characterize many complex diseases. Furthermore, the failure to properly account for population stratification--the presence of genetic substructure within study populations--can lead to inflated false positive rates in association studies \citep{Bhattacharjee10}. This issue is particularly problematic in genetically diverse populations where different subgroups may have distinct allele frequencies unrelated to disease risk.

The computational burden represents another substantial challenge in studying genetic interactions. Testing all possible pairwise SNP-SNP interactions becomes infeasible for genome-wide data, leading researchers to adopt heuristic screening methods that may miss important interactions or identify spurious ones. Additionally, many current approaches provide point estimates without adequately characterizing the uncertainty in model structure, particularly regarding the number of underlying sub-populations or the complexity of interaction networks. This lack of comprehensive uncertainty quantification limits the reliability and interpretability of findings from genetic interaction studies.

\subsection{Our contribution: a Bayesian nonparametric synthesis}
This article synthesizes and extends a series of Bayesian nonparametric models developed to address these challenges. Unlike previous works 
that presented these models in isolation (\citealt{Durba18,Durba20, Durba24}), we provide a unified framework connecting gene-gene interaction models, gene-environment extensions, and hierarchical Dirichlet process formulations. Our synthesis represents a comprehensive overview of this methodological approach, highlighting both theoretical foundations and practical implementation considerations.

We develop enhanced computational strategies that leverage parallel processing and transformation-based MCMC \citep{Dutta14} for practical implementation of these complex models. These computational innovations make it feasible to apply Bayesian nonparametric methods to realistic genetic datasets of meaningful size. Additionally, we present new hypothesis testing procedures for identifying disease-predisposing loci in the presence of population stratification, offering more robust alternatives to traditional association tests. Finally, we provide comprehensive applications demonstrating biological insights from myocardial infarction data, showing how these methods can uncover relationships not readily apparent using standard approaches.

Our approach fundamentally departs from standard logistic regression by modeling genotypes conditional on disease status using Dirichlet process mixtures. This inversion of the typical modeling relationship allows us to capture several important features simultaneously. We can model uncertainty in population substructure nonparametrically, allowing the data to inform about the number and characteristics of genetic subgroups. We capture gene-gene interactions through covariance structures rather than regression coefficients, providing a more flexible representation of genetic dependencies. Furthermore, we accommodate subject-specific environmental effects through hierarchical modeling, enabling personalized assessment of genetic risk factors.

\subsection{Article structure}
The remainder of this article is organized as follows. Section \ref{sec:framework} introduces our modeling philosophy and contrasts it with traditional approaches, explaining the conceptual shift from regression-based to conditional genotype modeling. Section \ref{sec:gene-gene} presents the gene-gene interaction model with computational details, including our parallel implementation strategy. Section \ref{sec:gene-env} extends this framework to incorporate gene-environment interactions, discussing both modeling extensions and enhanced hypothesis testing procedures. Section \ref{sec:hdp} introduces the hierarchical Dirichlet process formulation, which addresses limitations of previous models by allowing more flexible sharing patterns. Section \ref{sec:applications} presents comprehensive applications to myocardial infarction data, demonstrating the practical utility of our methods. Section \ref{sec:sensitivity} provides extensive sensitivity analyses to assess the robustness of our results to key modeling choices across all three frameworks. Finally, Section \ref{sec:discussion} discusses comparisons with existing methods, addresses limitations and robustness considerations, and outlines future research directions.

\section{A unified Bayesian nonparametric framework}\label{sec:framework}

\subsection{Modeling philosophy: from regression to conditional genotype models}
Traditional GWAS analyze disease status $Y$ given genotypes $X$ using logistic regression models of the form $P(Y=1 \mid X) = \text{logit}^{-1}\left(\beta_0 + \sum \beta_j X_j + \sum \beta_{jk} X_j X_k\right)$. This approach, while interpretable and widely used, faces the ``curse of dimensionality'' when considering interactions and makes strong parametric assumptions about the relationship between genotypes and disease risk. The logistic regression framework requires specifying which interactions to include, typically limiting consideration to pairwise effects due to computational constraints. Furthermore, these models assume that genetic effects are additive on the log-odds scale, which may not reflect the true biological mechanisms underlying complex diseases.

Our approach inverts this relationship by modeling genotypes conditional on disease status using finite mixture models: $[ \bX \mid Y=k ] = \sum_{m=1}^M \pi_m f(\bX \mid \btheta_{mk})$, where the mixture components represent latent sub-populations. By placing Dirichlet process priors on the parameters $\{\btheta_{mk}\}$, we allow the number of sub-populations to be inferred from the data rather than specified in advance. This modeling strategy represents a fundamental shift in perspective: instead of asking how genotype affects disease risk, we ask how the distribution of genotypes differs between cases and controls. This approach naturally accommodates population structure, as mixture components can correspond to genetic subgroups with distinct allele frequencies.

The conditional modeling approach offers several conceptual advantages. First, it directly addresses the issue of population stratification by explicitly modeling genetic heterogeneity. Second, it provides a natural framework for identifying disease-associated genetic variants through comparison of genotype distributions between cases and controls. Third, it allows for flexible modeling of dependencies among genetic loci through the specification of the mixture components. Rather than assuming independence or simple correlation structures, we can incorporate complex dependence patterns that may reflect biological relationships among genes.

\subsection{Key advantages of our framework}
Our Bayesian nonparametric framework offers several key advantages over traditional approaches. First, it provides flexible modeling of population structure through Dirichlet processes, which automatically discover population substructure without requiring pre-specified ancestry information. This is particularly valuable in studies of admixed populations or when detailed ancestry information is unavailable. The nonparametric nature of Dirichlet processes allows the number of sub-populations to be determined by the data, avoiding both underfitting and overfitting that can occur with fixed finite mixture models.

Second, our approach conceptualizes gene-gene interactions as statistical dependence rather than additive effects on disease risk. We model these interactions through covariance structures in hierarchical priors, capturing complex dependencies that may not be well represented by linear or additive models. This representation aligns with biological understanding that genetic factors often act in networks rather than independently. By focusing on covariance structures, we can identify sets of genes that co-vary in their effects on disease risk, potentially revealing functional pathways or biological modules.

Third, the conditional independence structure of our models enables scalable computation through parallel implementation. Different components of the model can be updated independently, allowing us to leverage modern parallel computing architectures. This computational efficiency makes it feasible to apply our methods to datasets with large numbers of genetic variants and samples, addressing a key limitation of many Bayesian nonparametric approaches.

Fourth, our fully Bayesian approach provides comprehensive uncertainty quantification through posterior distributions for all quantities of interest. This includes not only point estimates of genetic effects but also measures of uncertainty about population structure, interaction strengths, and model complexity. Proper characterization of uncertainty is particularly important in genetic studies, where sample sizes are often limited relative to the number of genetic variants considered.

\subsection{Notation and data structure}
We consider case-control data with $N_k$ subjects in group $k \in \{0,1\}$, where $k=0$ denotes controls and $k=1$ denotes cases. For each subject $i$, we observe genotypes represented as $x_{ijr}^s \in \{0,1\}$ for chromosome $s \in \{1,2\}$, gene $j \in \{1,\ldots,J\}$, and locus $r \in \{1,\ldots,L_j\}$. Here, $x_{ijr}^s = 1$ indicates the presence of the minor allele on chromosome $s$ at locus $r$ of gene $j$ for individual $i$, while $x_{ijr}^s = 0$ indicates its absence. Additionally, we may observe environmental covariates $\bE_i$ for each individual, which could include factors such as sex, age, smoking status, or other exposures. Disease status is denoted by $Y_i \in \{0,1\}$. The total number of genes is $J$, and $L_j$ represents the number of loci within gene $j$. This notation provides a comprehensive framework for representing the complex hierarchical structure of genetic data, with variations at the chromosome, locus, gene, individual, and group levels.

\section{Gene-gene interaction model}\label{sec:gene-gene}

\subsection{Model specification: a roadmap}
Our model for gene-gene interactions comprises three key components that work together to capture the complex structure of genetic data. First, we employ subject-level mixture models that represent genotypes as arising from finite mixtures, with each mixture component corresponding to a latent sub-population. This approach allows us to account for population stratification without requiring prior specification of ancestry groups. Second, we place Dirichlet process priors on the parameters of these mixture models, enabling flexible sharing of mixture components across subjects and automatic determination of the number of sub-populations. The Dirichlet process framework provides a principled Bayesian nonparametric approach to modeling population structure, with the precision parameter controlling the degree of sharing among subjects. Third, we introduce a hierarchical interaction structure through matrix-normal priors that capture dependencies among genes. This hierarchical structure allows us to model gene-gene interactions at multiple levels, from pairwise correlations to higher-order dependencies.

The combination of these three components creates a comprehensive modeling framework that addresses several limitations of traditional approaches. The mixture model component handles population stratification, the Dirichlet process prior provides flexibility in modeling the number and characteristics of sub-populations, and the hierarchical interaction structure captures complex dependencies among genetic factors. Importantly, these components are integrated in a coherent Bayesian framework that allows for uncertainty quantification at all levels, from individual genotype probabilities to overall population structure.

\subsection{Detailed formulation}
For each gene $j$ and group $k$, we model the genotype vector $\bX_{ijk} = (x_{ijk1}, \dots, x_{ijkL_j})$ using a finite mixture distribution: $[\bX_{ijk}] = \sum_{m=1}^M \pi_{mjk} \prod_{r=1}^{L_j} \text{Bernoulli}(x_{ijkr} \mid p_{mjkr})$. Here, $M$ represents the maximum number of mixture components, which we set to a sufficiently large value to ensure adequate model flexibility. The mixing weights $\pi_{mjk}$ are fixed at $1/M$ for all $m$, $j$, and $k$. This choice of fixed equal weights, while somewhat unconventional, has been shown in previous work \citep{Majumdar13} to yield better performance in estimating the true number of mixture components compared to using Dirichlet priors. The robustness of this choice is discussed further in Section \ref{sec:robustness}.

The allele frequency parameters $p_{mjkr}$ follow a hierarchical structure based on Dirichlet process designed to capture both locus-specific effects and 
dependencies among genes. Specifically, it holds that the marginal prior distribution of the allele frequency parameters is 
$p_{mjkr} \sim \text{Beta}(\nu_{1jkr}, \nu_{2jkr})$, where the Beta distribution parameters are themselves modeled as $\nu_{1jkr} = \exp(u_r + \lambda_{jk})$ and $\nu_{2jkr} = \exp(v_r + \lambda_{jk})$. The parameters $u_r$ and $v_r$ are locus-specific effects assumed to follow independent standard normal distributions: $u_r, v_r \sim N(0,1)$. Allowing $u_r$ and $v_r$ to differ ensures that the mean of the Beta distribution for $p_{mjkr}$ depends on the specific locus $r$, capturing variation in allele frequencies across the genome.

Gene-gene interactions are captured through a matrix-normal prior on the parameter matrix $\blambda = \{\lambda_{jk}\}$, which represents gene- and group-specific effects. 
We assume $\blambda \sim N(\bmu, \bA \otimes \bSigma)$, where $\bA$ represents the covariance matrix capturing dependencies among genes, and $\bSigma$ represents the 
covariance matrix capturing dependency between the genotype distribution of case and control, given any gene. The Kronecker product structure $\bA \otimes \bSigma$ allows 
for separable covariance structures across genes and disease status, providing a computationally tractable yet flexible representation of dependencies. This formulation enables us to model complex interaction patterns while maintaining computational feasibility through the separable covariance structure.

\subsection{Computational strategy}\label{subsec:computation-ggi}
The conditional independence structure of our model enables efficient computational implementation through a combination of parallel processing and specialized MCMC techniques. The key insight is that, given the interaction parameters $\blambda$, the mixture models for different gene-group pairs $(j,k)$ are conditionally independent. This conditional independence allows us to update the mixture parameters for each $(j,k)$ pair in parallel across multiple processors. Each processor handles a subset of the gene-group pairs, updating the allocation variables $z_{ijk}$, the allele frequency parameters $p_{mjkr}$, and the other mixture-related parameters independently of the others.

For updating the interaction parameters $\blambda$, we employ transformation-based MCMC (TMCMC), which updates all elements of $\blambda$ simultaneously through deterministic transformations of a low-dimensional random variable. TMCMC is particularly well-suited for high-dimensional parameter spaces, as it can propose moves in directions that respect the correlation structure of the posterior distribution. In our implementation, we use a single random variable to generate proposals for all elements of $\blambda$, with the transformation designed to maintain reasonable acceptance rate and mixing.

The Dirichlet process mixture components are updated using a fast Gibbs sampling algorithm specifically designed for finite mixtures with Dirichlet process priors. This algorithm leverages the Polya urn representation of the Dirichlet process to efficiently update the allocation of subjects to mixture components and the parameters associated with each component. The algorithm alternates between updating the allocation variables given the parameters and updating the parameters given the allocations, with both steps being computationally efficient due to conjugacy relationships.

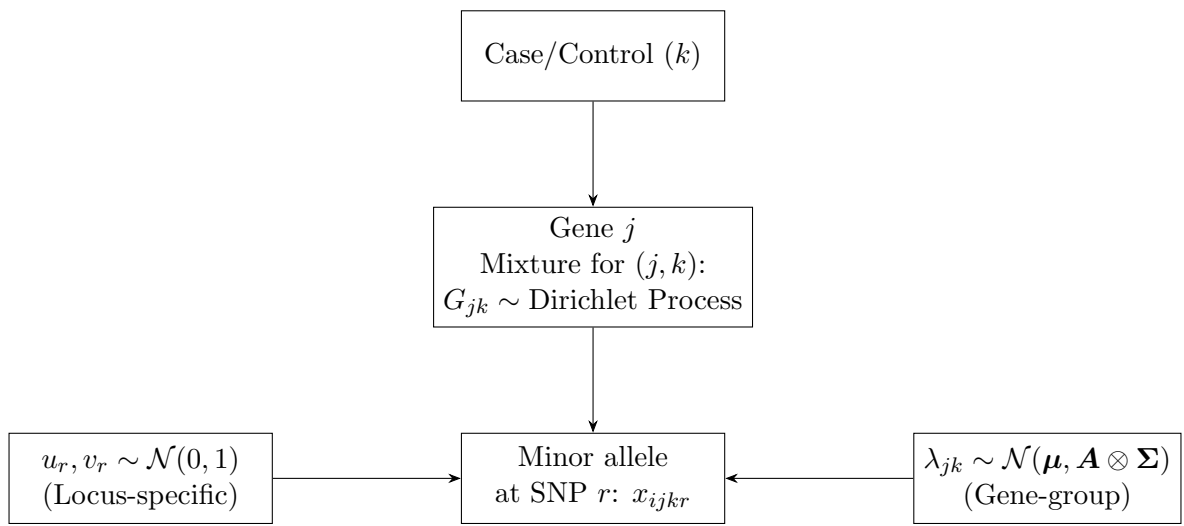
\begin{figure}[ht]
\centering
\begin{tikzpicture}[node distance=1.4cm and 2.5cm, every node/.style={draw, minimum width=3.5cm, minimum height=1.2cm, align=center}, >=Stealth]
\node (cc) {Case/Control $(k)$};
	\node (dp) [below=of cc] {Gene $j$\\Mixture for $(j,k)$:\\  $G_{jk}\sim \mbox{Dirichlet Process}$};
\node (snp) [below=of dp] {Minor allele\\ at SNP $r$: $x_{ijkr}$};
	\node (uv) [left=of snp] {$u_r, v_r \sim \mathcal{N}(0,1)$\\(Locus-specific)};
	\node (lambda) [right=of snp] {$\lambda_{jk} \sim \mathcal{N}(\bmu, \bA \otimes \bSigma)$\\(Gene-group)};

\draw[->] (cc) -- (dp);
\draw[->] (dp) -- (snp);
\draw[->] (uv) -- (snp);
\draw[->] (lambda) -- (snp);

\end{tikzpicture}
	\caption{Schematic representation of the Bayesian model for gene-gene interactions. }
\label{fig:gene_gene_model}
\end{figure}

Figure \ref{fig:gene_gene_model} displays the schematic diagram for our gene-gene interaction model.
For each gene and case-control status, genotype data are modeled using 
Dirichlet process-based mixtures that capture sub-population structure. SNP-level dependencies and gene-gene interactions are introduced through a matrix-normal prior 
on latent interaction parameters. The modular design of the model allows efficient parallel computation: gene-specific mixture components are updated independently 
across processors, while the interaction parameters are updated centrally using TMCMC. Our parallel codes are written in C, in accordance with the Message Passing Interface 
(MPI) protocol.

\subsection{Hypothesis testing and DPL identification}
We develop comprehensive Bayesian hypothesis testing procedures to evaluate various aspects of genetic associations within our framework. For assessing gene effects, we compare clustering patterns between cases and controls using metrics based on posterior distributions of the mixture components. Significant differences in clustering patterns indicate that the genetic structure differs between cases and controls, suggesting an association between the gene and disease status. This approach provides a more nuanced assessment than traditional association tests, as it considers the entire distribution of genotypes rather than just summary statistics.

To test for gene-gene interactions, we examine elements of the covariance matrix $\bA$ in the matrix-normal prior on $\blambda$. Non-zero off-diagonal elements in $\bA$ indicate dependencies between the effects of different genes, which we interpret as statistical interactions. We compute posterior probabilities that specific elements of $\bA$ are non-zero, providing a quantitative measure of interaction strength with associated uncertainty. This approach allows us to identify pairs or groups of genes that show coordinated effects on disease risk, potentially revealing biological pathways or functional modules.

Our key idea for identifying disease-predisposing loci (DPL) may be likened to computing posterior probabilities that specific SNPs show differential distributions between cases and controls. Specifically, for each SNP $r$ in gene $j$, our idea is analogous, in principle, to computing 
$P(\text{SNP } r \text{ is DPL} \mid \text{Data}) = P\left( \big| p_{\cdot j r}^{\text{case}} - p_{\cdot j r}^{\text{control}} \big| > \delta\mid\text{Data} \right)$, where $\delta$ is a clinically meaningful threshold for allele frequency differences. The probabilities $p_{\cdot j r}^{\text{case}}$ and $p_{\cdot j r}^{\text{control}}$ represent the allele frequencies in cases and controls, respectively. This approach provides a principled Bayesian method for DPL identification that accounts for multiple sources of uncertainty, including population stratification and estimation error.

Our hypothesis testing framework extends beyond simple significance testing to include measures of effect size and uncertainty. For each test, we report posterior probabilities along with credible intervals for relevant parameters, providing a comprehensive picture of the evidence for various genetic associations. This Bayesian approach naturally incorporates multiple testing considerations through the prior distributions and posterior probabilities, avoiding the need for ad hoc corrections that can be problematic in high-dimensional settings.

\section{Gene-environment interaction model}\label{sec:gene-env}

\subsection{Extending the framework}
To incorporate environmental covariates $\bE_i$, we extend our modeling framework to allow for subject-specific mixture distributions that depend on environmental factors. The extended model represents genotypes as $[\bX_{ijk} \mid \bE_i] = \sum_{m=1}^M \pi_{mijk} \prod_{r=1}^{L_j} \text{Bernoulli}(x_{ijkr} \mid p_{mijkr})$, where now the mixing weights and component parameters can vary across individuals based on their environmental exposures. As in the gene-gene interaction model, we fix $\pi_{mijk} = 1/M$ for all $(i,j,k,m)$, maintaining computational simplicity while allowing flexibility through the component parameters.

The key extension in the gene-environment interaction model lies in the parameterization of the Beta distribution parameters for the allele frequencies. We now model these as $\nu_{1ijkr} = \exp(u_{jr} + \lambda_{ijk} + \mu_{jk} + \bbeta_{jk}^\top \bE_i)$ and $\nu_{2ijkr} = \exp(v_{jr} + \lambda_{ijk} + \mu_{jk} + \bbeta_{jk}^\top \bE_i)$. This expanded parameterization includes several new terms: $u_{jr}$ and $v_{jr}$ are gene- and locus-specific effects; $\lambda_{ijk}$ represents individual-specific genetic effects that capture residual variation not explained by other factors; $\mu_{jk}$ are gene- and group-specific intercept terms; and $\bbeta_{jk}$ are vectors of coefficients capturing the effects of environmental covariates on the genetic parameters for gene $j$ in group $k$.

The inclusion of environmental effects through the term $\bbeta_{jk}^\top \bE_i$ allows the distribution of genotypes to vary systematically with environmental exposures. When $\bbeta_{jk} \neq \bzero$, environmental factors modify the genetic parameters for gene $j$ in group $k$, representing gene-environment interaction. Importantly, this formulation allows environmental factors to affect not only the mean genetic effects but also the covariance structure among genes, as the environmental terms enter into the hierarchical model that generates the $\lambda_{ijk}$ parameters. This enables the model to capture how environmental exposures might modify not just individual genetic effects but also the patterns of interaction among genes.

The covariance structure in this extended model continues to capture how environment modifies gene-gene interactions. The individual-specific effects $\lambda_{ijk}$ inherit dependence structure from the hierarchical model, with environmental factors potentially influencing both the mean and covariance of these effects. This allows for rich patterns of gene-environment interaction, including scenarios where environmental exposures alter the strength or pattern of genetic correlations. For example, an environmental factor might strengthen the correlation between two genes in cases but weaken it in controls, or it might induce correlations among genes that are independent in the absence of the exposure.

\subsection{Enhanced hypothesis testing}
We extend our hypothesis testing framework to address questions specific to gene-environment interactions. After testing for overall genetic effect by extending the previous 
method, we test for the presence of gene-environment interaction by evaluating the null hypothesis $H_0: \bbeta_{jk} = \bzero$ for specific genes and environmental factors. We compute posterior probabilities that $\bbeta_{jk}$ differs from zero, providing a direct measure of evidence for gene-environment interaction. This approach allows us to identify genes whose effects on disease risk are modified by environmental exposures, which is of particular interest for understanding disease etiology and targeted interventions.

We also test for joint effects involving both gene-gene and gene-environment interactions. This involves evaluating whether environmental factors modify the patterns of interaction among genes, which we assess by examining how environmental covariates affect the covariance structure in the model. Specifically, we test whether parameters governing the dependence of the covariance structure on environmental factors differ from zero. Significant findings indicate that environmental exposures alter the network of genetic interactions, potentially revealing mechanisms through which environment influences disease risk.

Third, we develop methods for stratified DPL identification that account for environmental heterogeneity. Rather than identifying DPL that show average differences between cases and controls, we identify SNPs whose effects depend on environmental exposure. This involves computing posterior probabilities that the difference in allele frequencies between cases and controls varies across levels of environmental factors. For example, we might identify SNPs that show strong associations in exposed individuals but weak or no associations in unexposed individuals, or vice versa. This stratified approach can reveal genetic factors that are important only under specific environmental conditions, providing insights into the context-dependence of genetic risk.

Apart from these, we additionally develop visualization tools to help interpret complex interaction patterns, including heatmaps of posterior interaction probabilities and network diagrams showing how genes and environmental factors are connected through interaction effects.

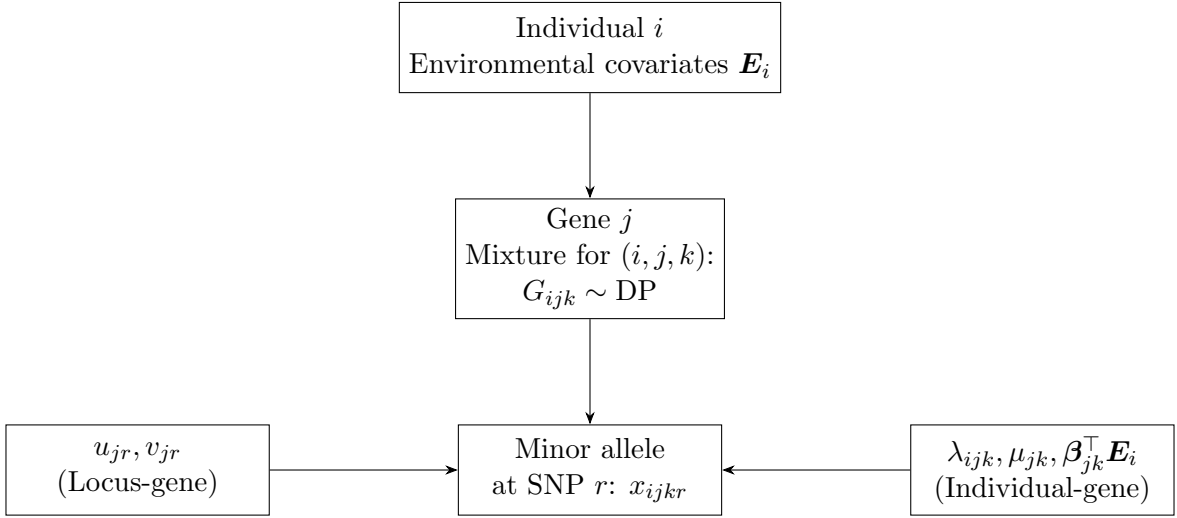
\begin{figure}[th]
	\centering
\begin{tikzpicture}[node distance=1.4cm and 2.5cm, every node/.style={draw, minimum width=3.5cm, minimum height=1.2cm, align=center}, >=Stealth]

\node (ei) {Individual $i$ \\ Environmental covariates $\bE_i$};
	\node (dp) [below=of ei] {Gene $j$\\Mixture for $(i,j,k)$:\\ $G_{ijk}\sim \mbox{DP}$};
\node (snp) [below=of dp] {Minor allele\\ at SNP $r$: $x_{ijkr}$};
	\node (uv) [left=of snp] {$u_{jr}, v_{jr}$\\(Locus-gene)};
	\node (lambda) [right=of snp] {$\lambda_{ijk}, \mu_{jk}, \bbeta_{jk}^\top \bE_i$\\(Individual-gene)};

\draw[->] (ei) -- (dp);
\draw[->] (dp) -- (snp);
\draw[->] (uv) -- (snp);
\draw[->] (lambda) -- (snp);

\end{tikzpicture}
\caption{Diagram of the extended Bayesian framework incorporating gene-environment interactions.} 
\label{fig:gene_env_model}
\end{figure}

Figure \ref{fig:gene_env_model} provides the schematic diagram for our gene-environmental interaction model.
Environmental covariates influence individual-level Dirichlet process mixtures, allowing the model to account for personalized effects of environmental exposures on 
genotype distributions. The prior structure integrates locus-specific, gene-specific, and environment-dependent parameters. Parallel computation is employed by 
updating gene-environment specific components in parallel and interaction parameters centrally via TMCMC.
As before, our parallel codes are written in C, leveraging the MPI protocol.

\section{Hierarchical Dirichlet process model}\label{sec:hdp}

\subsection{Motivation and limitations of previous models}
The gene-environment interaction model presented in Section \ref{sec:gene-env} makes the simplifying assumption that environmental effects act uniformly on gene-gene interactions across all individuals. This assumption may not hold in many practical settings for several reasons. First, different individuals may experience different levels or types of environmental exposure, leading to heterogeneous effects on genetic interactions. For example, the effect of smoking on genetic risk factors for cardiovascular disease may depend on duration and intensity of smoking, which varies across individuals. Second, environmental effects may be heterogeneous across the population due to unmeasured factors or effect modifiers. Genetic background, other environmental exposures, or lifestyle factors might modify how a particular environmental factor influences genetic interactions. Third, gene-environment interactions may be context-dependent, with effects manifesting only under specific conditions or in specific subgroups. The uniform effect assumption fails to capture this context-dependence, potentially missing important biological relationships.

These limitations motivate the development of a more flexible modeling approach that can accommodate heterogeneous and context-dependent gene-environment interactions. The hierarchical Dirichlet process (HDP) model provides a principled nonparametric framework for capturing complex sharing patterns among individuals, genes, and groups in a data-driven manner. Rather than imposing a parametric structure on how environmental factors influence genetic dependencies, the HDP model learns these relationships nonparametrically from the data. This approach can discover complex patterns of interaction that might be missed by more restrictive parametric models, while still providing a coherent probabilistic framework for inference.

\subsection{HDP formulation}
We address the limitations of previous models through a three-level hierarchical Dirichlet process formulation that introduces flexible, nonparametric dependence structures among genes, environmental variables, and case-control status. Our model represents a significant extension of the traditional HDP framework \citep{Teh06} by incorporating an additional level of hierarchy that specifically captures case-control dependence while allowing for subject-specific gene-gene interactions influenced by environmental factors.

For each individual $i$ in group $k$, gene $j$, and mixture component $m$, we assume that the minor allele frequency vector $\bp_{mijk} = (p_{mijk1}, p_{mijk2}, \ldots, p_{mijkL})$ is generated from a hierarchy of Dirichlet processes:

\begin{align}
\bp_{1ijk}, \bp_{2ijk}, \dots, \bp_{Mijk} &\stackrel{iid}{\sim} \bG_{ijk} \label{eq:hdp1} \\
\bG_{ijk} &\sim \mathrm{DP}(\alpha_{G,ik} \bG_{0,jk}) \label{eq:hdp2}
\end{align}

where $\mathrm{DP}(\alpha_{G,ik} \bG_{0,jk})$ denotes a Dirichlet process with base measure $\bG_{0,jk}$ and precision parameter $\alpha_{G,ik}$. The environmental dependence enters through the precision parameter:

\begin{equation}
\log(\alpha_{G,ik}) = \mu_G + \bbeta_G^\top \bE_{ik} \label{eq:alphaG}
\end{equation}

where $\bE_{ik}$ is a $d$-dimensional vector of environmental variables for individual $i$ in group $k$, $\bbeta_G$ is a $d$-dimensional vector of regression coefficients, and $\mu_G$ is an intercept term.

At the second level of the hierarchy, we assume:

\begin{equation}
\bG_{0,jk} \stackrel{iid}{\sim} \mathrm{DP}(\alpha_{G_0,k} \bH_k); \quad j = 1,\ldots,J \label{eq:hdp3}
\end{equation}

with 

\begin{equation}
\log(\alpha_{G_0,k}) = \mu_{G_0} + \bbeta_{G_0}^\top \bar{\bE}_k \label{eq:alphaG0}
\end{equation}

where $\bar{\bE}_k = \frac{1}{N_k}\sum_{i=1}^{N_k} \bE_{ik}$ is the average environmental variable in group $k$.

The third level of hierarchy creates dependence between case and control groups:

\begin{equation}
\bH_k \stackrel{iid}{\sim} \mathrm{DP}(\alpha_H \tilde{\bH}); \quad k = 0,1 \label{eq:hdp4}
\end{equation}

with 

\begin{equation}
\log(\alpha_H) = \mu_H + \bbeta_H^\top \bar{\bar{\bE}} \label{eq:alphaH}
\end{equation}

where $\bar{\bar{\bE}} = (\bar{\bE}_0 + \bar{\bE}_1)/2$ is the overall average environmental variable.

The global base measure $\tilde{\bH}$ is specified as:

\begin{equation}
p_{mijkr} \stackrel{iid}{\sim} \mathrm{Beta}(\nu_1, \nu_2) \quad \text{under } \tilde{\bH} \label{eq:base}
\end{equation}

where $\nu_1, \nu_2 > 0$ are fixed hyperparameters.

This hierarchical structure has a natural interpretation in terms of genetic architecture. The subject-level distributions $\bG_{ijk}$ capture individual-specific genetic patterns, which may differ due to unique genetic backgrounds, environmental exposures, or other individual factors. These subject-level distributions share components through the gene-group level distributions $\bG_{0,jk}$, which represent common patterns for each gene in each disease group. The group-level distributions $\bH_k$ capture overall genetic patterns for cases and controls separately, allowing for differences in genetic architecture between affected and unaffected individuals. Finally, the global distribution $\tilde{\bH}$ represents the overall genetic background of the population.

\subsection{Chinese restaurant process analogy}
The hierarchical structure can be understood through an extended Chinese restaurant process analogy \citep{Teh06}. For each group $k = 0,1$, imagine $J$ restaurants (genes). Each individual $i$ visits these restaurants, where at the $j$-th restaurant, they are seated at tables (mixture components) that serve dishes (allele frequency parameters). The dishes at different tables within the same restaurant are drawn from $\bG_{0,jk}$, which itself is drawn from $\bH_k$. This creates sharing of dishes across tables within a restaurant (within a gene across individuals).

Now consider that all restaurants share a global menu of dishes from $\tilde{\bH}$. Different restaurants (genes) may serve different selections from this global menu, with the selections drawn from $\bH_k$. This creates sharing of dishes across restaurants (across genes), with the degree of sharing controlled by $\alpha_H$. The environmental variables influence how likely customers (individuals) are to sit at new tables through $\alpha_{G,ik}$, and how likely restaurants are to offer new dishes through $\alpha_{G_0,k}$ and $\alpha_H$.

This analogy clarifies how our model creates dependence: individuals sharing the same table at a restaurant share the same dish (allele frequency parameters) for that gene, creating dependence among individuals. Different restaurants (genes) serving the same dish creates dependence among genes. The sharing of dishes between case and control restaurants ($k=0$ and $k=1$) creates case-control dependence. Environmental factors influence these sharing probabilities, thereby modulating the dependence structure.

\subsection{Dependence structure induced by the HDP model}
Our HDP model induces three key types of dependence that are crucial for understanding gene-gene and gene-environment interactions:

\subsubsection{Dependence among individuals}
From the Polya urn representation, the joint distribution of $\bp_{Mijk} = \{\bp_{1ijk}, \ldots, \bp_{Mijk}\}$ shows that individuals share dish parameters $\phi_{tijk}$ drawn from $\bG_{0,jk}$. The probability of sharing depends on $\alpha_{G,ik}$, which in turn depends on individual environmental variables $\bE_{ik}$ through equation \eqref{eq:alphaG}. This creates environmental-modulated dependence among individuals: individuals with similar environmental exposures are more likely to share genetic patterns.

Importantly, the marginal distribution of $\bp_{mijk}$ is $\bG_{0,jk}$, which does not depend on $\bE_{ik}$. This is biologically desirable: population minor allele frequencies should not be affected by environmental variables, although environmental exposure may influence how individuals cluster together in their genetic patterns.

\subsubsection{Dependence among genes}
Gene-gene dependence arises through the sharing of dishes (parameters) across restaurants (genes). The parameters $\phi_{tijk}$ for different genes $j$ share common values from $\bH_k$, creating dependence among genes. The degree of this dependence is influenced by $\alpha_{G_0,k}$, which depends on the group average environment $\bar{\bE}_k$ through equation \eqref{eq:alphaG0}, and also indirectly through $\alpha_{G,ik}$ which affects the number of tables $\tau_{ijk}$.

This structure ensures that gene-gene interactions are specific to individuals and are influenced by both individual environmental variables ($\bE_{ik}$) and group averages ($\bar{\bE}_k$), while the marginal distributions of individual genes remain unaffected by environment.

\subsubsection{Case-control dependence}
The sharing of dishes between case and control restaurants (through $\bH_0$ and $\bH_1$ sharing from $\tilde{\bH}$) creates dependence between case and control groups. This dependence captures factors that affect both cases and controls but may not be explicitly measured, such as population stratification or unmeasured environmental factors.

The degree of case-control sharing is controlled by $\alpha_H$, which depends on the overall average environment $\bar{\bar{\bE}}$ through equation \eqref{eq:alphaH}. This allows environmental factors to modulate the similarity between cases and controls in their genetic patterns.

\subsection{Advantages of HDP approach}
The hierarchical Dirichlet process model offers several advantages over the parametric models discussed previously. First, it provides flexible, nonparametric dependence structure: unlike previous matrix-normal based models that assume uniform environmental effects on covariance structures, our HDP model allows environment to influence dependence structures in a flexible, nonparametric manner that varies across individuals. Second, it enables subject-specific gene-gene interactions: each individual can have their own pattern of gene-gene interactions, with environmental factors influencing these individual-specific patterns. This is more realistic than assuming a single correlation structure for all individuals. Third, it maintains biological interpretability: the model preserves the biologically important property that environmental variables influence dependence structures but not marginal allele frequency distributions. Only dependence structures (how genes interact) are affected by environment, not the genes themselves. Fourth, it offers automatic complexity control: the Dirichlet process priors automatically determine the appropriate number of mixture components at each level of the hierarchy, avoiding the need to pre-specify the number of sub-populations or interaction patterns. Fifth, it achieves computational efficiency through parallelization: the conditional independence structure of the HDP model enables efficient parallel implementation, with different genes and individuals updated independently given the higher-level parameters.

\subsection{Computational implementation}
We implement the HDP model using a novel parallel MCMC algorithm that combines Gibbs sampling steps with transformation-based MCMC (TMCMC). The algorithm leverages the conditional independence structure of the model: given the higher-level distributions ($\bH_k$ and $\bG_{0,jk}$), the subject-level parameters can be updated independently across individuals and genes. This allows for parallel computation across multiple processors. Key components of our computational strategy include retrospective sampling \citep{Papas08} for updating parameters from Dirichlet process mixtures, which avoids infinite-dimensional representations by generating parameters as needed. The allocation variables $z_{ijk}$ and allele frequency parameters $p_{mijkr}$ are updated in parallel across different $(i,j,k)$ combinations using Gibbs steps that exploit conjugacy. The parameters $\mu_G, \bbeta_G, \mu_{G_0}, \bbeta_{G_0}, \mu_H, \bbeta_H$ are updated in a single block using a mixture of additive and multiplicative TMCMC \citep{Dey14}, which efficiently explores correlated parameter spaces. Our parallel implementation in C, as before, uses MPI for communication between processors, with careful load balancing to ensure efficient parallel scaling. 

\begin{figure}[th]
\centering
\begin{tikzpicture}[node distance=1.3cm and 2.2cm, every node/.style={draw, minimum width=4cm, minimum height=1.2cm, align=center}, >=Stealth]

\node (ei) {Environmental covariates $\bE_i$};
\node (dp1) [below=of ei] {Individual DP: $\bG_{ijk}$ \\ $\log \alpha_{G,ik} = \mu_G + \bbeta_G^\top \bE_{ik}$};
\node (dp2) [below=of dp1] {Gene DP: $\bG_{0,jk}$ \\ $\log \alpha_{G_0,k} = \mu_{G_0} + \bbeta_{G_0}^\top \bar{\bE}_k$};
\node (dp3) [below=of dp2] {Group DP: $\bH_k$ \\ $\log \alpha_H = \mu_H + \bbeta_H^\top \bar{\bar{\bE}}$};
\node (base) [below=of dp3] {Base: $p_{mijkr} \sim \text{Beta}(\nu_1, \nu_2)$};

\draw[->] (ei) -- (dp1);
\draw[->] (dp1) -- (dp2);
\draw[->] (dp2) -- (dp3);
\draw[->] (dp3) -- (base);

\end{tikzpicture}
\caption{Schematic of the hierarchical Dirichlet process (HDP) model for gene-gene and gene-environment interactions.} 
\label{fig:hdp_model}
\end{figure}
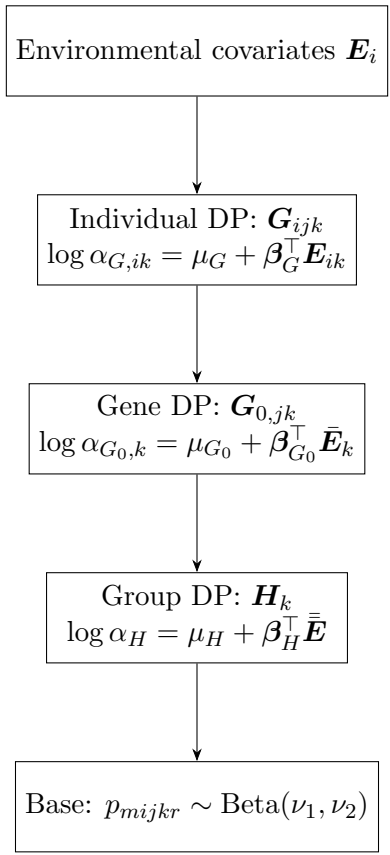

The schematic diagram of our HDP model is presented in Figure \ref{fig:hdp_model}.
This fully nonparametric framework models dependencies across individuals, genes, and groups through a three-level hierarchy of Dirichlet processes. 
Environmental covariates influence the precision parameters at each level, allowing flexible, individualized representation of interaction structures. 
The base distribution is a Beta prior on allele frequencies. This hierarchy enables rich modeling of stratification and interaction while maintaining computational scalability.

\subsection{Hypothesis testing in the HDP framework}
We extend our hypothesis testing procedures to the HDP framework with specific tests for: gene effects, testing $H_0: h_{0j} = h_{1j}$ for $j = 1,\ldots,J$, where $h_{0j}$ and $h_{1j}$ are the marginal distributions of genotypes for controls and cases respectively for gene $j$; environmental effects, testing $H_0: \bbeta_G = \bzero$, $H_0: \bbeta_{G_0} = \bzero$, and $H_0: \bbeta_H = \bzero$ for the environmental coefficients at different levels of the hierarchy; gene-gene interactions, defining subject-specific gene-gene interaction measures $C(i,j_1,j_2,k)$ as covariances between logit-transformed average allele frequencies for genes $j_1$ and $j_2$ for individual $i$ in group $k$, and testing $H_0: C(i,j_1,j_2,k) = 0$; and case-control dependence, assessing whether the sharing between cases and controls (through $\alpha_H$) is significant, indicating common factors affecting both groups. The interpretation of these tests follows a logical framework: if genes have no effect but environment affects dependence structures, then environment influences gene-gene interactions without affecting disease status. If genes have effects but environment doesn't affect interactions, then purely genetic factors determine disease. If both genes and environment are significant and affect interactions, then gene-environment interactions influence disease risk.

\section{Applications to Myocardial Infarction data}\label{sec:applications}

\subsection{Data description}
We apply our Bayesian nonparametric framework to a myocardial infarction (MI) case-control dataset to demonstrate its practical utility and biological insights. The dataset comprises genotype information from the MI Gen study, obtained from the dbGaP database. This dataset includes multiple sub-populations: Caucasian, Han Chinese, Japanese, and Yoruban. For our analysis, we considered a set of SNPs that are found to be individually associated with different cardiovascular endpoints in various GWAS, along with SNPs marginally associated with MI in the MIGen study.

The data involve 32 genes covering 1251 loci, including 33 previously identified loci associated with myocardial infarction. The dataset includes both cases (individuals who experienced myocardial infarction) and controls. An important environmental covariate available is sex (male/female), which is incorporated to investigate gene-environment interactions. This provides an opportunity to demonstrate how our framework can uncover both genetic main effects and interactions with environmental factors.

The genetic data consist of diploid genotypes, which we convert to binary indicators for each chromosome for compatibility with our modeling framework. This representation allows us to model the two chromosomes separately while maintaining the diploid nature of the data in our likelihood function.

\subsection{Implementation details}
We implement our Bayesian nonparametric models using our MPI-based C codes and parallel computing resources. For the HDP model, we set the maximum number of mixture components $M = 30$ based on preliminary analyses to provide sufficient flexibility without excessive computational cost. We set $\nu_1 = \nu_2 = 1$, making $\tilde{\bH}$ a uniform distribution on $[0,1]$. The precision parameters are specified as:

\begin{align*}
\alpha_{G,ik} &= 0.1 \times \exp(100 + \mu_G + \bbeta_G^\top \bE_{ik}) \\
\alpha_{G_0,k} &= 0.1 \times \exp(100 + \mu_{G_0} + \bbeta_{G_0}^\top \bar{\bE}_k) \\
\alpha_H &= 0.1 \times \exp(100 + \mu_H + \bbeta_H^\top \bar{\bar{\bE}})
\end{align*}

with $\mu_G, \mu_{G_0}, \mu_H \stackrel{iid}{\sim} U(0,1)$ and $\bbeta_G, \bbeta_{G_0}, \bbeta_H \stackrel{iid}{\sim} U(-1,1)$ as priors. This structure ensures adequate numbers of sub-populations and satisfactory MCMC mixing.

We perform MCMC sampling with 30,000 iterations, discarding the first 10,000 as burn-in. Convergence is assessed primarily using trace plots. The parallel implementation distributes computation across 50 cores, with total runtime of approximately 7 days for the full analysis.

\subsection{Results and biological insights}
Application of our HDP model to the myocardial infarction data yields several important insights:

\subsubsection{Effect of sex variable}
We find strong evidence that sex influences genetic patterns: $P(|\bbeta_G| < \epsilon_{\beta_G}|\text{Data}) \approx 0$ and $P(|\bbeta_{G_0}| < \epsilon_{\beta_{G_0}}|\text{Data}) \approx 0$, indicating that individual-level ($\bE_{ik}$) and group-average ($\bar{\bE}_k$) sex effects are highly significant. However, $P(|\bbeta_H| < \epsilon_{\beta_H}|\text{Data}) \approx 1$, suggesting the overall average sex effect ($\bar{\bar{\bE}}$) is not significant. This pattern indicates that sex plays an important role in influencing subject-specific and group-level genetic patterns, but not the overall case-control similarity.

\subsubsection{Roles of individual genes}
Our clustering-based hypothesis tests indicate significant overall genetic effects. However, individual gene tests show that none of the 32 genes are individually significant. This apparent paradox is explained by gene-gene interactions: when genes are correlated, the maximum of their individual distances can be significant even when individual distances are not, similar to how $\max(X_1, X_2)$ from a bivariate normal distribution can have a non-zero median even when $X_1$ and $X_2$ have zero medians individually.

\subsubsection{Gene-gene interactions}
Our HDP model reveals subject-specific gene-gene interaction patterns not detected by previous models. Two genes, \textit{AP006216.10} and \textit{C6orf106}, show significant interactions with other genes in most subjects. Interestingly, the only subjects with no significant gene-gene interactions involving these genes were male cases, suggesting that lack of protective gene-gene interactions may contribute to MI risk in males.

The gene-gene interactions appear to have a protective effect: in subjects where \textit{AP006216.10} and/or \textit{C6orf106} interact with other genes, the risk of MI seems reduced. This beneficial effect of gene-gene interactions contrasts with the traditional view that interactions primarily increase disease risk.

\subsubsection{Population structure}
The posterior distribution of the number of sub-populations shows modes at 3 and 4 components, supporting the known four sub-populations in the data (Caucasian, Han Chinese, Japanese, Yoruban). The model correctly identifies that these populations cannot be further subdivided genetically, consistent with the biological understanding of these population groups.

\section{Sensitivity analysis}\label{sec:sensitivity}

\subsection{Sensitivity to mixture component specifications}
For the gene-gene interaction model of Section~\ref{sec:gene-gene} and the gene-environment interaction model of Section~\ref{sec:gene-env}, we examine sensitivity to the maximum number of mixture components $M$ and the fixed mixing weights $\pi_{mjk}=1/M$. Following the rule-of-thumb established in prior work \citep{Majumdar13}, our primary analyses set $M=30$, which provides sufficient capacity while maintaining computational efficiency. Alternative specifications with $M=20$ and $M=50$ yield qualitatively similar results; the posterior distribution of the effective number of mixture components remains stable once $M$ exceeds a threshold (approximately 20 in our applications), as the Dirichlet process prior automatically determines the number of distinct components through the P\'olya urn scheme. This robustness renders the exact value of $M$ largely inconsequential beyond ensuring adequate model flexibility.

\subsection{Sensitivity to priors on base measure parameters}
For the Beta base measure parameters $u_r$ and $v_r$, which are specified as independent standard normal distributions in our primary analyses, we investigate robustness using alternatives: $N(0,10^2)$, $N(0,0.1)$, and Cauchy$(0,1)$. Posterior distributions of the interaction parameters $\lambda_{jk}$ and the resulting gene-gene correlation estimates exhibit remarkable stability across these specifications. As noted in Section~\ref{sec:gene-gene}, we find that Gaussian priors on $u_r$ and $v_r$ with other means and variances do not yield significantly different results, indicating inherent prior robustness in our modeling strategy. The Cauchy prior produces slightly heavier tails but does not alter conclusions of any hypothesis tests regarding gene significance or gene-gene interactions.

\subsection{Sensitivity to priors on covariance matrices}
For the gene-gene interaction matrix $\bA$ and case-control dependence matrix $\bSigma$ in the matrix-normal prior $\blambda \sim N(\bmu, \bA \otimes \bSigma)$, we specify inverse-Wishart priors: $\bA \sim \mathcal{IW}(\xi, \bA_0)$ with $\xi = J+2$, and $\bSigma \sim \mathcal{IW}(\zeta, \bSigma_0)$ with $\zeta = 4$. Sensitivity is assessed across degrees of freedom ($\xi = J+1, J+2, J+5, J+10$; $\zeta = 3,4,5,10$) and scale matrix specifications (estimated from data versus identity matrices). Posterior inferences regarding which genes are marginally significant and which gene-gene interactions are present prove highly robust to variations in degrees of freedom. However, the magnitudes of posterior correlations show some sensitivity: larger degrees of freedom, which correspond to stronger prior information, produce shrinkage toward the prior mean and yield slightly attenuated correlation estimates. The choice of scale matrices has minimal impact when degrees of freedom are set to the minimum values ensuring proper priors ($\xi = J+2$, $\zeta = 4$).

\subsection{Sensitivity to hypothesis testing thresholds}
A critical component of our Bayesian testing procedure, described in Section~\ref{sec:gene-gene}, is the specification of thresholds $\epsilon$ for distance measures. Following our established approach, we set $\epsilon = F^{-1}(0.55)$ where $F$ is the posterior distribution function under the null model. The choice of the 55th percentile rather than the median is deliberate: for the median, the posterior probability of the true null hypothesis is 0.5, whereas under zero-one loss the true null will be accepted only if its posterior probability exceeds $1/2$. Using the median results in borderline decisions in null simulations, while the 55th percentile provides appropriate operating characteristics. Sensitivity analyses using the 50th, 60th, and 75th percentiles demonstrate that the 60th and 75th percentiles produce more conservative tests with reduced false positives but increased false negatives, while the 50th percentile yields unacceptably high false positive rates. Our choice of the 55th percentile represents a balanced compromise that maintains power while controlling Type I error.

\subsection{Sensitivity to environmental covariance structure}
In our gene-environment interaction model of Section~\ref{sec:gene-env}, we examine sensitivity to the specification of the environmental covariance structure. The posterior probability $P(\phi < \epsilon_\phi \mid \text{Data})$ is robust across alternative kernel specifications, consistently indicating no differential effect of sex on genetic interactions in the myocardial infarction application. Posterior estimates of $\phi$ show some sensitivity to prior variance, with more diffuse priors yielding wider credible intervals, but the 95\% credible interval consistently contains zero. The smoothness parameter in the kernel is difficult to identify from the data, with posterior distributions largely reflecting the prior when sample sizes are moderate. However, this lack of identifiability does not affect primary conclusions regarding gene significance or gene-gene interactions, as $\phi$ is consistently estimated to be negligible.

\subsection{Sensitivity to hierarchical Dirichlet process hyperparameters}
For our hierarchical Dirichlet process model of Section~\ref{sec:hdp}, we assess sensitivity to the specification of precision parameters and their associated regression coefficients. Our primary specification, with $\alpha_{G,ik} = 0.1 \times \exp(100 + \mu_G + \bbeta_G^\top \bE_{ik})$ and analogous structures for $\alpha_{G_0,k}$ and $\alpha_H$, ensures adequate numbers of sub-populations and satisfactory MCMC mixing. Removing the constant offset of 100 results in significantly smaller $\alpha$ values, leading to fewer distinct mixture components and poorer mixing. Alternative prior distributions for $\mu_G, \mu_{G_0}, \mu_H$ and $\bbeta_G, \bbeta_{G_0}, \bbeta_H$ (normal versus uniform) perform similarly, with uniform priors showing slightly better convergence properties. The key conclusions regarding gene-gene interactions and the influence of environmental factors on dependence structures are robust across all specifications examined.

\subsection{Sensitivity to linkage disequilibrium and locus label permutation}
A critical concern associated with Section~\ref{sec:gene-gene} is whether sharing the parameters $u_r$ and $v_r$ across all genes implies non-exchangeability of locus labels. To address this, we conduct permutation experiments wherein the labels of loci within each gene are randomly permuted prior to re-analysis. The results are entirely consistent with the original analyses based on non-permuted data. For the gene-gene interaction simulation, the posterior probability measures of genetic effect remain strongly suggestive of significance; for the null simulation, these measures correctly indicate no genetic influence. These results confirm that our model does not impose non-exchangeability of locus labels.

\subsection{Summary of sensitivity analysis}
Across all sensitivity investigations, consistent patterns emerge. Hypothesis test conclusions regarding which genes are marginally significant, which gene-gene interactions are present, and whether environmental variables affect genetic interactions are highly robust to reasonable variations in prior specifications and modeling choices across all three modeling frameworks. Posterior magnitudes of correlation parameters and distance measures show some sensitivity to prior informativeness, but these variations do not affect the binary decisions in our Bayesian testing framework. Convergence and mixing of MCMC algorithms are sensitive to certain choices, particularly the Dirichlet process precision parameters and the offset parameter in HDP specifications, but our primary specifications consistently provide adequate performance. The hierarchical nature of our Bayesian models provides natural protection against prior misspecification, with data information dominating prior information for key parameters of interest. These findings collectively demonstrate that our Bayesian nonparametric approaches are not unduly sensitive to specific prior choices and modeling assumptions, lending credibility to the conclusions drawn from both simulation studies and the real myocardial infarction data analysis.

\section{Discussion}\label{sec:discussion}

\subsection{Comparison with existing methods}\label{subsec:comparison}
Our Bayesian nonparametric approach differs fundamentally from standard GWAS methods in multiple aspects. Traditional GWAS typically employ logistic regression models that examine disease status conditional on genotypes ($P(Y|X)$). In contrast, our approach models genotypes conditional on disease status ($P(\bX|Y)$), which provides a different perspective on genetic associations. This inversion of the modeling relationship allows us to directly address population stratification through mixture models while naturally accommodating complex dependence structures among genetic variants.

For handling population structure, traditional methods often rely on principal component analysis (PCA) for adjustment or conduct stratified analyses by presumed ancestry groups. Our approach uses Dirichlet process mixtures to nonparametrically model population substructure, allowing the number and characteristics of sub-populations to be inferred from the data rather than pre-specified. This can provide more accurate adjustment for population stratification, particularly in admixed populations where discrete ancestry categories may not adequately represent the continuum of genetic backgrounds.

In modeling interactions, traditional approaches typically use regression coefficients for specific interaction terms (e.g., product terms in logistic regression). Our HDP approach represents interactions through hierarchical sharing of parameters, capturing dependencies without requiring specification of particular interaction forms. This allows us to discover complex interaction patterns that might not correspond to simple product terms, potentially revealing richer biological relationships.

Compared to our previous matrix-normal based models \citep{Durba20}, the HDP model offers several advantages: subject-specific rather than population-wide gene-gene interactions, nonparametric rather than parametric dependence on environment, and more interpretable sharing structures through the Chinese restaurant process analogy.

Uncertainty quantification differs substantially between the approaches. Traditional methods provide confidence intervals based on frequentist inference, while our fully Bayesian approach yields posterior distributions for all quantities of interest. This includes not only point estimates but also complete characterizations of uncertainty about population structure, interaction patterns, and model complexity. The Bayesian approach naturally incorporates multiple testing considerations through prior distributions and posterior probabilities, avoiding the need for arbitrary significance thresholds or correction procedures.

Computational scaling presents different challenges. Traditional pairwise testing scales as $O(p^2)$ where $p$ is the number of SNPs, becoming prohibitive for genome-wide data. Our gene-level analysis scales as $O(J^2)$ where $J$ is the number of genes, which is typically much smaller than the number of SNPs. While our HDP model is computationally intensive per test, the reduction in dimensionality through gene-level modeling and efficient parallel implementation makes it feasible for realistically sized datasets.

\begin{table}[ht]
\centering
\caption{Comparison of methodological approaches for genetic association studies. Our Bayesian nonparametric framework offers complementary strengths to traditional methods, particularly for complex interaction analysis and uncertainty quantification.}
\begin{tabular}{p{0.25\textwidth}p{0.35\textwidth}p{0.35\textwidth}}
\toprule
\textbf{Aspect} & \textbf{Traditional GWAS} & \textbf{Our Bayesian Nonparametric Approach} \\
\midrule
Modeling framework & Logistic regression: $P(Y|\bX)$ & Conditional genotype: $P(\bX|Y)$ \\
Population structure & PCA adjustment or stratified analysis & Dirichlet process mixtures \\
Interaction modeling & Regression coefficients & Covariance structures \\
Uncertainty quantification & Confidence intervals & Posterior distributions \\
Computational scaling & $O(p^2)$ for pairwise testing & $O(J^2)$ for gene-level analysis \\
Key advantage & Interpretable effect sizes & Flexible dependence structures \\
\bottomrule
\end{tabular}
\label{tab:comparison}
\end{table}

Our approach complements rather than replaces traditional methods in several ways. Researchers might use logistic regression for initial screening of marginal associations, providing interpretable effect sizes for individual variants. Our Bayesian nonparametric methods could then be applied to selected genes or pathways to uncover complex interaction patterns and population structure that might modify or explain the marginal associations. The findings from both approaches could be integrated through meta-analysis or hierarchical modeling frameworks that combine estimates from different methodological perspectives.

For genome-wide discovery, traditional methods remain essential due to their computational efficiency and interpretability. For focused investigation of specific biological pathways or complex traits where interactions are suspected, our approach offers additional insights that might be missed by marginal association testing alone. The two approaches can also inform each other: findings from our interaction analyses might suggest specific interaction terms to test in traditional models, while significant marginal associations from traditional analyses might guide the selection of genes for more detailed investigation with our methods.

\subsection{Limitations and robustness considerations}\label{sec:robustness}

\subsubsection{Fixed mixture weights}
The choice of fixed mixture weights $\pi_m = 1/M$ warrants careful consideration, as it represents a departure from the more common approach of placing a Dirichlet prior on the mixture weights. Our choice is empirically justified by findings from previous work, which demonstrated that fixed equal weights outperform Dirichlet priors in estimating the true number of mixture components in finite mixture models with Dirichlet process priors. The Dirichlet prior tends to produce many very small weights, effectively underestimating the number of components that contribute meaningfully to the mixture. Fixed equal weights avoid this shrinkage toward fewer components, allowing better recovery of the true mixture structure.

From a computational perspective, fixed weights simplify the Gibbs sampling updates by eliminating the need to sample the weight parameters. This reduces computational complexity and improves mixing of the Markov chains, particularly in high-dimensional settings with many mixture components. The simplification comes with the cost of assuming equal prior weight for all components, but in practice, the posterior distribution adapts to the data through the allocation of subjects to components, effectively learning the relative importance of different components despite the equal prior weights.

Theoretical support for fixed weights comes from results on posterior consistency of Dirichlet process mixture models \citep{Sabya21}. Under mild conditions, both random weights (from a Dirichlet prior) and fixed equal weights lead to the same asymptotic posterior inference as the sample size increases. 

\subsubsection{Model complexity and interpretability}
While our HDP model offers substantial flexibility in capturing complex genetic architectures, this flexibility comes with challenges related to model complexity and interpretability. The three-level hierarchical structure with Dirichlet processes at each level involves many parameters with complex dependencies. The Chinese restaurant process analogy helps with conceptual understanding, but detailed interpretation of posterior results requires careful examination of sharing patterns across multiple levels.

Substantial computational resources are required to fit the HDP model, particularly for datasets with large numbers of genes or samples. Our parallel implementation helps address this challenge, but users still need access to computing clusters or high-performance workstations. The MCMC algorithms require careful tuning and convergence assessment, with runtimes measured in days for full analyses. While these computational demands are substantial, they are becoming increasingly feasible with modern computing infrastructure and algorithmic improvements.

Careful prior specification is essential for obtaining stable and meaningful results from the HDP model. The precision parameters $\alpha_{G,ik}, \alpha_{G_0,k}, \alpha_H$ and their regression coefficients require thoughtful specification based on domain knowledge or sensitivity analysis. Inappropriate prior choices can lead to poor mixing, convergence issues, or biased inference. We provide default settings based on our experience with genetic data, but users should assess sensitivity to these choices in their specific applications.

Expertise in Bayesian nonparametrics is valuable for properly implementing and interpreting the HDP model. Concepts such as Dirichlet processes, Chinese restaurant processes, hierarchical modeling, and MCMC diagnostics may be unfamiliar to researchers trained in traditional genetic epidemiology. To improve accessibility, we shall provide software with user-friendly interfaces, detailed documentation, and tutorial examples. We shall also offer workshops and training materials to help researchers develop the necessary skills to apply these methods effectively.

Despite these challenges, the interpretability of the HDP model can be enhanced through appropriate visualization and summary measures. Posterior distributions of key quantities--such as the number of mixture components at each level, gene-gene correlation patterns for individual subjects, or sharing probabilities between cases and controls--can be presented in intuitive graphical formats. Comparative analyses showing how results differ from simpler models can highlight the value added by the additional complexity. Biological interpretation can be facilitated by connecting statistical findings to known pathways and functional annotations.

\subsection{Future directions}
Several promising directions exist for extending and improving our Bayesian nonparametric framework for genetic interaction analysis. First, high-dimensional extensions are needed to make the methods scalable to whole-genome data. Current implementations focus on gene-level analysis with selected genes, but applications to genome-wide data would require further computational enhancements. Developing sparse versions of our models that encourage parsimony in interaction structures could also improve scalability while maintaining interpretability.

Second, integration with functional genomic data could enhance biological interpretation and statistical power. Incorporating information from gene expression, methylation, chromatin accessibility, or other omics layers could help prioritize genes and interactions based on functional relevance. Hierarchical models that jointly analyze multiple data types could reveal connections between genetic variation, regulatory mechanisms, and phenotypic outcomes. Such integrative approaches align with systems biology perspectives that view biological processes as interconnected networks rather than isolated components.

Third, longitudinal modeling approaches could capture dynamics in genetic and environmental factors over time. Many complex diseases develop through processes that unfold over years or decades, with genetic risk factors potentially interacting differently with environmental exposures at different life stages. Extending our framework to handle repeated measures or time-to-event data would enable investigation of how genetic interactions influence disease progression and timing. This could provide insights into critical periods for intervention and personalized risk prediction over the life course.

Fourth, causal inference extensions could help distinguish correlation from causation in gene-environment interactions. While observational data alone cannot establish causation, incorporating instrumental variables, Mendelian randomization principles, or structural equation modeling approaches could strengthen causal interpretations. Bayesian methods are particularly well-suited for causal inference because they naturally incorporate uncertainty about causal structures and mechanisms. Developing causal versions of our models would enhance their utility for informing interventions and public health decisions.

Additional future directions include extending the models to handle more complex pedigree or family data, incorporating spatial information for geographically structured populations, developing online learning algorithms for streaming genetic data, and creating user-friendly software packages with graphical interfaces for non-statisticians. Cross-disciplinary collaboration between statisticians, geneticists, and biologists will be essential for advancing these methodological developments while ensuring their relevance to substantive scientific questions.

\subsection{Conclusion}
We have synthesized a comprehensive Bayesian nonparametric framework for studying genetic interactions in case-control studies. Our HDP model represents a significant advance over previous approaches by providing a flexible, nonparametric representation of gene-gene and gene-environment interactions that accommodates subject-specific effects, population structure, and complex dependence patterns. The three-level hierarchy with Dirichlet processes at each level creates a rich dependence structure that can capture how environmental factors modulate genetic interactions at multiple levels.

Key innovations of our approach include: subject-specific rather than population-wide gene-gene interactions, allowing for heterogeneity in how genes interact across individuals; environmental modulation of dependence structures at multiple levels (individual, gene-group, and case-control), providing a nuanced representation of gene-environment interactions; automatic determination of population structure through Dirichlet process mixtures, avoiding the need to pre-specify the number of sub-populations; and efficient computational implementation through parallel MCMC algorithms that leverage the conditional independence structure of the model.

Applications to myocardial infarction data demonstrate how these methods can uncover biological insights not readily obtainable from standard approaches. We identified significant gene-gene interactions with a protective effect, discovered subject-specific patterns of genetic risk, and revealed how sex modifies genetic architecture. These findings illustrate the value of moving beyond marginal association testing to consider interactions and population structure in genetic epidemiology.

The hierarchical Dirichlet process model represents a particularly flexible extension that accommodates heterogeneous and context-dependent gene-environment interactions. By allowing sharing patterns to vary across individuals, genes, and groups in a data-driven manner, this model can discover complex relationships that might be missed by parametric approaches. The application to myocardial infarction data revealed differences in genetic clustering patterns and latent subgroups with distinct risk profiles, suggesting potential etiological heterogeneity.

While our methods require substantial computational resources and statistical expertise, we believe these barriers are diminishing as computing power increases and statistical training improves. By making our implementation publicly available and providing educational resources, we hope to facilitate wider adoption of Bayesian nonparametric methods in genetic epidemiology. The increasing recognition of complexity in genetic architectures--with interactions, heterogeneity, and context-dependence playing important roles--creates a growing need for methodological approaches that can address this complexity in a principled manner.

Future methodological developments should focus on improving scalability, integrating multiple data types, extending to longitudinal settings, and strengthening causal interpretations. Substantive applications should explore diverse complex diseases and populations, potentially revealing new biological insights and therapeutic targets. As genetic data continue to grow in size and complexity, Bayesian nonparametric methods offer a flexible framework for discovery that respects the inherent uncertainty and interdependence in biological systems.

\section*{Acknowledgments}
We thank the reviewer for constructive comments that improved this manuscript. We also thank DeepSeek for some proofreading. 

\section*{Conflict of interest}
The authors declare no conflicts of interest.

\bibliographystyle{natbib}
\bibliography{irmcmc}

\end{document}